\documentclass[a4paper,10pt,twocolumn]{article}
\usepackage{bm}
\usepackage{graphicx}
\usepackage{ascmac}
\usepackage[margin=20truemm]{geometry}
\usepackage{authblk}
\usepackage{url}

\author[1]{Soichiro Ueda \thanks{ueda@inet.media.kyoto-u.ac.jp}}
\author[2]{Ai Nozaki\thanks{nozaki@hal.ipc.i.u-tokyo.ac.jp}}
\author[1]{Daisuke Kotani\thanks{kotani@media.kyoto-u.ac.jp}}
\author[1]{Yasuo Okabe\thanks{okabe@media.kyoto-u.ac.jp}}
\affil[1]{University of Tokyo \\ Tokyo, Japan}
\affil[2]{Kyoto University \\ Kyoto, Japan}

\date{}

\begin{document}

\title{Mewz: Lightweight Execution Environment for WebAssembly with High Isolation and Portability using Unikernels}
\maketitle

\begin{abstract}
  Cloud computing requires isolation and portability for workloads.
  Cloud vendors must isolate each user's resources from others to prevent them from attacking other users or the whole system.
  Users may want to move their applications to different environments, for instance other cloud, on-premise servers, or edge devices.
  Virtual machines (VMs) and containers are widely used to achieve these requirements.
  However, there are two problems with combined use of VMs and containers.
  First, container images depend on host operating systems and CPU architectures.
  Users need to manage different container images for each platform to run the same codes on different OSes and ISAs.
  Second, performance is degraded by the overheads of both VMs and containers.
  Previous researches have solved each of these problems separately, but no solution solves both problems simultaneously.
  Therefore, execution environments of applications on cloud are required to be more lightweight and portable while ensuring isolation is required.
  We propose a new system that combines WebAssembly (Wasm) and unikernels.
  Wasm is a portable binary format, so it can be run on any host operating systems and architectures.
  Unikernels are kernels statically linked with applications, which reduces the overhead of guest kernel.
  In this approach, users deploy applications as a Wasm binary and it runs as a unikernel on cloud.
  To realize this system, we propose a mechanism to convert a Wasm binary into a unikernel image with the Wasm AoT-compiled to native code.
  We developed a unikernel with Wasm System Interface (WASI) API and an Ahead-of-Time (AoT) compiler that converts Wasm to native code.
  We evaluated the performance of the system by running a simple HTTP server compiled into Wasm and native code.
  The performance was improved by 30\% compared to running it with an existing Wasm runtime on Linux on a virtual machine.
\end{abstract}

\section{Introduction}
\label{sec:intro}

Cloud computing and container-based virtualization technologies are widely used in modern system development.
In cloud computing, multiple users share the same physical resources of a data center.
In such environments, it might be possible for users to attack other users by exploiting vulnerabilities in the system.
For example, malicious users can steal sensitive information from other users or destroy the whole cloud system.
To achieve multi tenacy, cloud providers must isolate each user's resources from others.
There are several isolation mechanisms, and the intensity of isolation depends on the type.

The most isolated mechanism is separating the physical machines.
Providing a physical machine for each user prevents interference between users.
This method, called bare-metal cloud\cite{baremetalcloud}, is not mainstream for some drawbacks such as difficulty in immediate resource allocation.

The second most isolated mechanism is virtual machine (VM) based isolation.
In this method a hypervisor runs on a physical machine and creates multiple VMs.
Each VM is isolated from others, and it has its own operating system.
This method is widely used nowadays.

The third most isolated mechanism is container-based isolation.
An container is an isolated environment that runs on a host operating system.
Since containers share the same kernel, they are more lightweight than VMs.
However, containers does not provide enough isolation for multi-tenancy\cite{gvisor}.

Despite the lack of isolation, containers are widely used in cloud computing for their portability.
Container images can be easily moved between different environments, such as development, testing, and production.
Containers are used in cloud environments with VMs to ensure isolation.

Combining containers and cloud computing entails two problems.
First, container images depend on host operating systems and CPU architectures.
To distribute container images to environments with different operating systems and architectures, developers must build images for each one.
Second, the virtualization overhead of both containers and VMs degrades performance.

There are existing solutions to each of these challenges.
For example, substitution of containers with WebAssembly is proposed to solve the portability problem\cite{wasmcloud}.
Binary translation is also proposed, which converts instructions of one architecture to another\cite{binary-translation}.
However, these solutions do not solve the overhead problem.
To reduce the overheads of guest kernel on VMs, previous researches proposed unikernels, which are specialized kernels for each application\cite{mirageos}.
There is also a work that implements a lightweight hypervisor with limited functionality to reduce the overhead of hypervisors\cite{firecracker}.
Conversely, these solutions do not enhance application portability.
No solution solves both problems.

We propose a new system where applications are distributed as WebAssembly and run them as unikernels on cloud.
WebAssembly is a portable binary format, so it can be run on any host operating systems and architectures.
Unikernels are kernels statically linked with applications.
This design enables applications to call kernel functions directly, reducing the overhead from guest OSes.
This system solves both the portability and virtualization overhead problems.
To implement this system, we need to link WebAssembly binaries with kernel codes.
However, the issue lies in converting a WebAssembly binary into a unikernel image, because WebAssembly cannot be simply linked with kernel codes.
Moreover, it is preferable to compile WebAssembly ahead of time (AoT) into native code for performance reasons.
We devise a new mechanism to do it by exploiting WebAssembly System Interface, which is the standardized API for WebAssembly to access system resources.
We combine an AoT-compiled WebAssembly binary and kernel codes that provide WebAssembly System Interface by symbol resolution.
We realized it by developing a unikernel that provides WebAssembly System Interface and an AoT compiler that converts WebAssembly to native code.

We evaluate our system with an HTTP server that distributes static files.
The result shows that our system executes Wasm applications with lower overhead than existing technologies.

The rest of this paper is organized as follows.
Section \ref{sec:background} summarizes the requirements for the execution environment of workloads on cloud and describes the problems of existing solutions.
Section \ref{sec:architecture} explains the architecture and implementation of the proposed system.
Section \ref{sec:performance} evaluates the performance of our system.
Section \ref{sec:related} describes related works.
Section \ref{sec:conclusion} concludes this paper.

\section{Background}
\label{sec:background}

There are two main requirements for cloud computing: isolation and portability.
\begin{description}
  \item[\texttt{isolation}:]
        Each workload must be sufficiently isolated to avoid interference between users\cite{SUBASHINI20111}.
        For example, it must be impossible for malicious users to steal data from other users' workloads or hijack their processes.
        Cloud providers must ensure kernel-level isolation between workloads of different tenants\cite{kernel-isolation}.

  \item[\texttt{portability}:]
        Cloud applications should be easily moved to different cloud environments or even to on-premise servers.
        Users have many choices for cloud providers, and they may want to change providers for cost or performance reasons\cite{di2016cloud}.
        Additionally, multi-cloud, in which multiple cloud vendors are used at the same time, is sometimes used to prevent vendor lock-in in recent years\cite{hong2019overview}.
\end{description}

To achieve the two requirements, VMs and containers are usually used.
Users deploy their applications as container images and cloud providers run them on VMs using hypervisors.
Containers must be used with VMs because containers do not provide isolation of host kernels unlike VMs.
However, there are two problems with this method.
This study focuses on solving these problems at the same time.

First, the portability of container images is restricted to the same host operating system and the same architecture.
Containers cannot run on different operating systems since they share the same kernel with the host
Container images also contain binaries compiled for a specific architecture, so they cannot be executed on different architectures.
This enforces container images to be made for each operating system and architecture.
It becomes a problem in environments such as cloud-edge continuum\cite{cloud-edge-continuum}, where the same applications run on both cloud and resource-restricted edge devices with different ISAs and OSes.
Applications must be distributed in such a way that they can run on any OSes and architectures.

Second, the overheads incurred by VMs and containers, respectively, occur simultaneously.
One of the overheads of virtual machines is incurred when applications access hardware\cite{vmoverhead}.
Without virtual machines, applications request the kernel to operate hardware via system calls, and the kernel accesses it directly.
Applications on a virtual machine issue system calls to the guest kernel, and the guest kernel operates virtual devices provided by the hypervisor.
In this case, the hypervisor emulates the behavior of the devices and accesses the actual hardware.
Thus, an application on a virtual machine goes through the guest kernel and the hypervisor before accessing the hardware, thus causing larger overhead than the case without a virtual machine.
Network isolation is one of the major overheads for containers\cite{suo2018analysis}.
In order to separate the container network from the host one, the container is assigned a virtual network interface created by the host.
This requires the container to have the host mediate packets when communicating with the outside world.
In this case, packet processing occurs twice: once at the virtual network interface of the container and once at the physical network interface of the host.
This results in larger packet processing time compared to the case without containers.
Consequently, using containers on a virtual machine for isolation and portability accepts both overheads, resulting in a significant performance degradation\cite{xu2013managing}.
Therefore, a more lightweight execution environment with both isolation and portability is required.

Previous researches have proposed solutions to each of these problems.
Our research aims to solve both problems at the same time, by combining two solutions.

\subsection{unikernels}

Unikernels\cite{unikernel} are a kind of lightweight kernel designed to run in cloud.
With unikernels, a single application is statically linked with the kernel into a single kernel image at build time.

In the traditional ways, applications run on a general-purpose operating system such as Linux or Windows.
These operating systems have many functions that are not necessary for the application.
In addtion, the kernel and the application are separated, so the application must issue system calls to the kernel to access hardware.

In contrast, since unikernels are designed to run only one application, they can be optimized for the application\cite{mirageos}.
Also, an application is combined with kernel into a single image.
Thus, the application can call kernel functions directly, which reduces the overhead of system calls.

However, since unikernel is built for an ISA, the kernel image of unikernel depends on the architecture.
Running the same application with unikernels on different architectures requires building a unikernel image for each architecture.
Therefore, unikernels do not solve the portability problem.

\subsection{Wasm}

WebAssembly\cite{WebAssemblyCoreSpecification1} (Wasm) is a portable binary format for executable code.
Wasm can be compiled from various programming languages, for example, C, C++, Rust, Go, Kotlin, and so on.
Wasm was originally designed to run on browsers, but in recent years, runtimes have appeared that run Wasm directly on operating systems.
For example, there are many implementations such as Wasmtime\cite{wasmtime}, WasmEdge\cite{wasmedge}, Wasmer\cite{wasmer}.
With such Wasm runtimes, Wasm is expected to be applied in various fields such as IoT\cite{edge}, serverless computing\cite{serverless}, and blockchain\cite{blockchain}.

Wasm has a specification of API called WebAssembly System Interface\cite{wasi} (WASI) that provides system resources to Wasm.
WASI allows Wasm to use resources such as file system and network.
In other words, WASI acts like a system call in Wasm. Unlike general system calls, WASI provides its interface in the form of functions.
WASI is defined only as a specification of its API, and its implementation is left to each runtime that uses WASI.

The OS/CPU architecture-independence of Wasm and WASI's ability to absorb platform differences make Wasm a good candidate for solving the portability problem.
Instead of distributing applications in the form of container images, applications may be compiled into Wasm and deployed.
This allows the distribution of a single Wasm binary regardless of OSes and architectures of target environments.

However, Wasm alone cannot provide the isolation required for cloud computing\cite{hyperlight}.
This is because Wasm shares the host kernel as well as containers.
If developers run Wasm in cloud, virtual machine isolation will be required.
This means that using Wasm as an alternative to containers does not solve the virtualization overhead issue.

\section{Architecture}
\label{sec:architecture}

This section proposes a new system that combines Wasm and unikernels to solve the problems of portability and overheads.

\begin{figure}
  \centering
  \includegraphics[width=\linewidth]{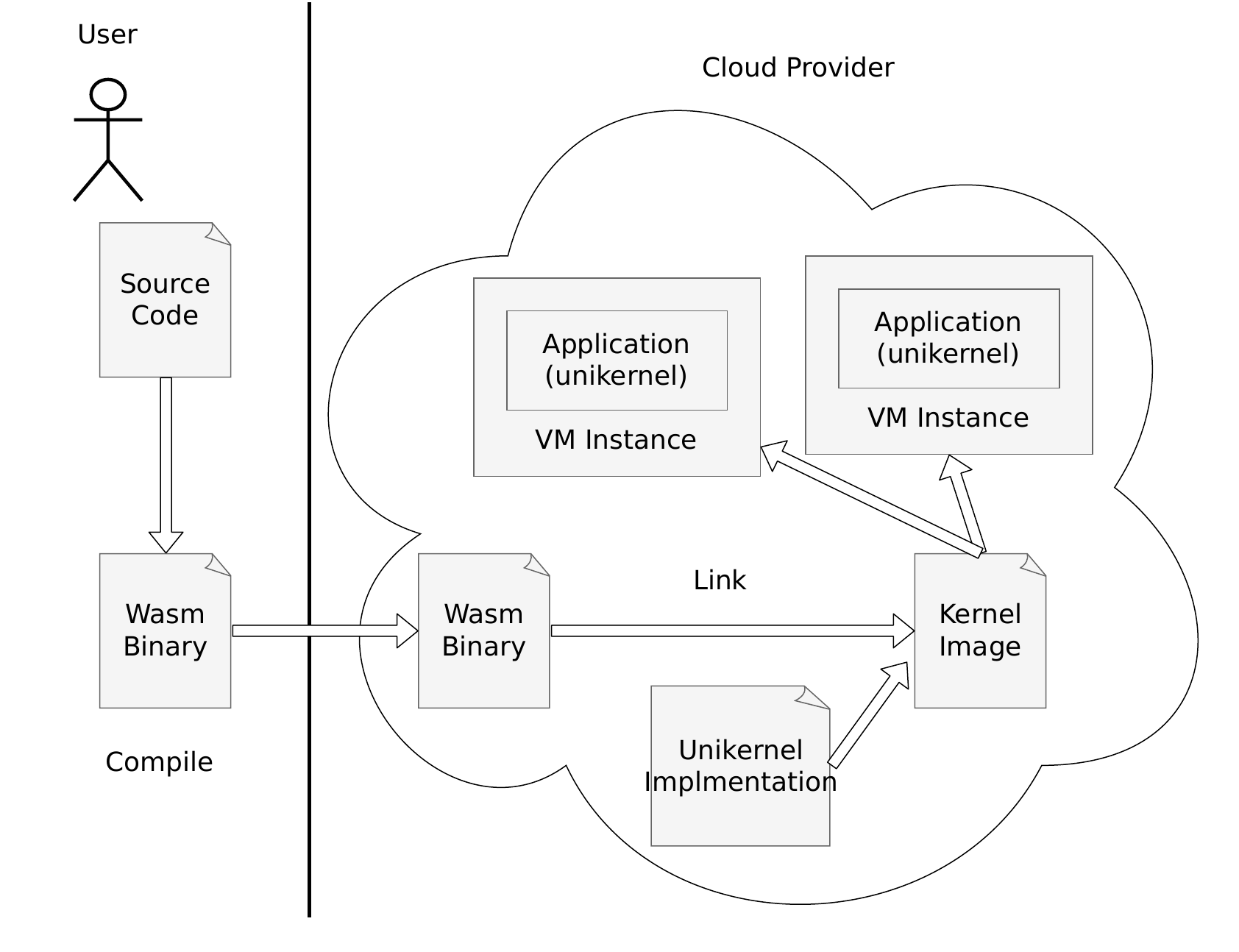}
  \caption{Architecture of the system}
  \label{fig:architecture}
\end{figure}

First, applications are compiled into Wasm binaries and distributed.
In cloud environments, Wasm binaries are linked into unikernel images.
The unikernel images are then run on VMs. The figure \ref{fig:architecture} shows the architecture of the system.

\begin{figure}
  \centering
  \includegraphics[width=\linewidth]{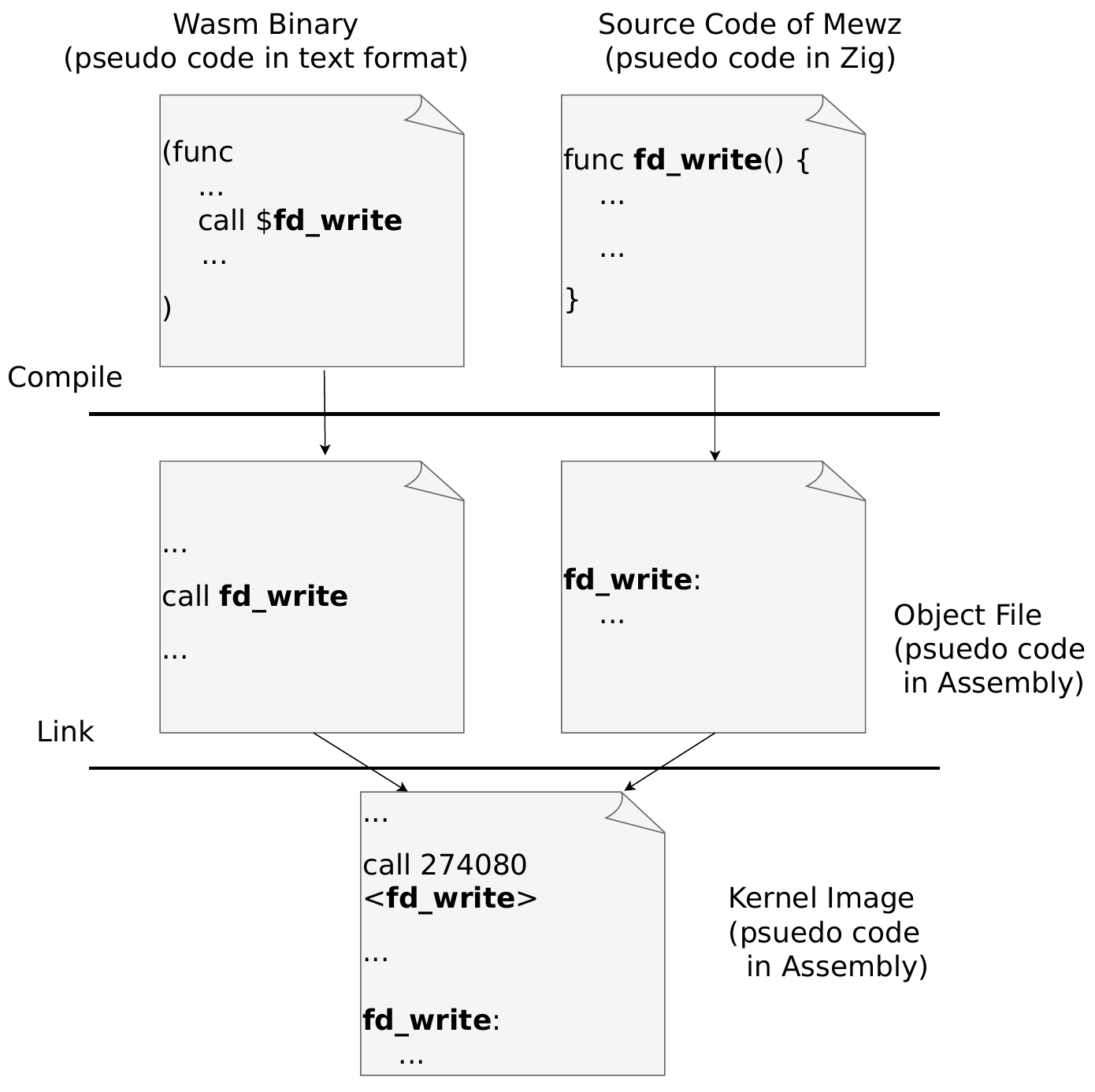}
  \caption{Linking process of Wasm and Mewz}
  \label{fig:linking}
\end{figure}

This architecture benefits from the lightweightness of unikernels while ensuring the portability of binaries.
With conventional unikernels, users must build a kernel image from source codes of applications, and distribute it to cloud.
In this way, users cannot deploy applications to environments with other architectures quickly and easily.
Some unikernels can execute Linux binaries\cite{hermitux}.
Even with these unikernels, the built binary is dependent on the OS and the architecture.
Thus, it is difficult to distribute portable binary enjoying the advantages of unikernels.
With this system, developers can run applications with low overheads on cloud while they can deploy the same binary to any environments, such as edge devices with other OSes or architectures.

To implement this system, we need to convert Wasm binaries into unikernel images.
But Wasm is a virtual ISA so it cannot be linked with kernel codes directly.
Additionally, it is desirable to AoT compile Wasm to native code for better performance.
Thus, it is a challenge to combine a Wasm binary and kernel codes into a unikernel image while converting Wasm to native codes.

The key to solving this challenge is WASI.
Since WASI is a set of functions, Wasm binaries have instructions that call WASI functions.
As shown in the figure \ref{fig:linking}, we can build a unikernel image containing a Wasm application.
First, we compile a Wasm binary into an object file with WASI function symbols unresolved.
Then, we link the object file with kernel codes that have WASI functions.
Thus, WASI acts as a boundary between an application and kernel codes, enabling Wasm to be linked with kernel codes.
No other portable binary format, such as JVM bytecode and LLVM IR, has this feature.
To do the above steps, we developed two softwares: Mewz and Wasker.

\subsection{Unikernel with WASI API}

We developed a unikernel, Mewz\cite{mewz}, that provides WASI API to Wasm binaries.
Mewz executes a Wasm application by statically linking with it at build time.
When a Wasm application linked with Mewz calls a WASI function, it jumps to the WASI function implemented in Mewz.
In this way, Mewz provides WASI API functions to Wasm and executes it.

Since Mewz is specialized for Wasm execution, its functions are provided only through WASI.
Mewz has only the functionality necessary for WASI implementation, which minimizes the functionality.
For example, since the current version of WASI does not have thread API, Mewz does not implement thread functionality.
As a result, Mewz does not have a thread scheduling function.
This shortage can be covered by scaling out the number of VMs.
It reduces the overhead of thread creation, switching, and scheduling.

\subsubsection{Memory Management}

Mewz has only one memory space for both the kernel and the application.

\begin{description}
  \item[\texttt{0x00000000 00100000 - 0x00007fff ffffffff}:]
        The text segment, the static data, and the stack are placed here.
        This range of address is mapped to the same address as physical memory.
  \item[\texttt{0xffff800 00000000 - 0xffff800 7fffffff}:]
        Wasm has a linear memory space, called Linear Memory.
        Wasm has instructions to access Linear Memory and to dynamically change the size of it.
        Linear Memory is placed in this area. \\
        Wasm regards Linear Memory as a memory space starting the address of 0.
        For this reason, Mewz need to prepare an exclusive memory region for Linear Memory.
        Mewz maps this area to physical memory according to the requested size of Linear Memory.
        Linear Memory is 32-bit address space, so it is used only up to 0xfff800 ffffffffff.
  \item[\texttt{0xffffc00 00000000 - 0xffffffff ffffffff}:]
        The heap of the kernel is placed here.
        The heap area is separated into this address range to facilitate the management of the heap.
        This area of address is also mapped to physical memory according to the needed size.
\end{description}

\subsubsection{Network}

In the version of WASI preview1\cite{wasi-p1}, the APIs that provide the networking functions of WASI are not yet complete.
For example, some APIs such as sending and receiving operations for sockets have been defined, but there are no APIs for creating sockets and setting IP addresses.

However, some of the existing Wasm runtimes extend WASI to provide networking capabilities on their own.
For example, WasmEdge has extended and added enough APIs to provide networking features such as socket creation and IP addressing.
Wasmer has also developed its own WASI extension API called WASIX\cite{wasix}, which is similar to POSIX and provides networking functions.

Mewz aims at a WASI preview1 compliant implementation, but at the same time, it requires an API that provides network functionality since it is intended to run on cloud.
We have implemented the network functionality in a way that is compatible with the existing Wasm runtime extension APIs.
Among the existing extension APIs, the WasmEdge extension can be used from various languages, so we implemented socket APIs compatible with WasmEdge.

The TCP/IP protocol stack is lwIP\cite{lwip}, an open source implementation for embedded systems.
The supported network device is virtio-net\cite{virtio-spec}.

\subsubsection{File System}

Mewz provides an on-memory, read-only file system.
It makes a tar archive of a specified directory on the host and includes it in the kernel image.
During the boot process, the tar archive is extracted into memory.
Mewz maintains the correspondence between the file names and the pointers to the metadata and the content.

Currently, Mewz only supports on-memory file system and does not support persistence of file system.
In the future, we plan to support mechanisms such as 9pfs\cite{9pfs} and virtiofs\cite{virtio-spec} to share the host file system with virtual machines to realize persistence.

\subsection{AoT Compiler of Wasm}

In order to statically link Mewz and a Wasm binary, we implemented Wasker\cite{wasker}, an AoT compiler that converts a Wasm binary to native code on a target CPU architecture.
Wasker compiles a Wasm binary into an object file, leaving WASI functions as unresolved symbols.
This means the object file converted from Wasm does not contain WASI implementations.
After AoT compilation by Wasker, the object file is linked with Mewz by symbol resolution, as shown in the figure \ref{fig:linking}.
Wasker uses LLVM as a backend for AoT compilation.

Wasker leaves two other functions besides the WASI functions as unresolved symbols to provide Linear Memory for Wasm.
They are a function to get the base address of Linear Memory and a function to change the size of Linear Memory.
When Wasm reads or writes to Linear Memory, it calculates the actual memory address to be accessed from the base address of Linear Memory.
These functions are also implemented by Mewz.

\section{Performance Evaluation}
\label{sec:performance}

For evaluation, we ran a simple HTTP server that distributes static files on the system.
We implemented the server in Rust and the source code can be compiled into both Wasm and native code.
We compared the performance of the following four environments.

\begin{description}
  \item[Mewz] \mbox{} \\
        We compiled the server into Wasm and ran it on Mewz.
  \item[WasmEdge] \mbox{} \\
        WasmEdge is a Wasm runtime that runs on Linux.
        WasmEdge has an AoT compilation feature for Wasm, which allows Wasm to be executed quickly.
        We compiled the server into Wasm and ran it on WasmEdge on Linux by AoT compilation.
        We used Ubuntu Server 22.04.3 LTS as the Linux distribution.
  \item[Nanos] \mbox{} \\
        Nanos\cite{nanos} is a unikernel that executes Linux application binaries.
        We compiled the server into native code targeting Linux and ran it on Nanos.
  \item[Linux] \mbox{} \\
        We compiled the server into native code targeting Linux and ran it on Linux.
        We used Ubuntu Server 22.04.3 LTS as the Linux distribution.
\end{description}

For benchmarking, we used an HTTP benchmarking tool called wrk\cite{wrk}.
wrk continuously sends HTTP requests to a specified URL and measures the throughput.
wrk keeps sending requests in parallel with a specified number of threads while maintaining a specified number of connections.
In this evaluation, 16 threads and 800 connections were loaded for 60 seconds.

We prepared two physical machines and wrk was executed on one of them
A KVM virtual machine was created on the other machine and the target was executed on it.
The two physical machines were connected with an L2 switch.
On the KVM host, a Linux Bridge was created and connected to the TAP device assigned to the virtual machine.

The benchmark machine was equipped with an AMD Ryzen 9 5900HX CPU, 32GB of memory, and a 2.5Gbps ethernet interface.
The KVM host machine had Intel Xeon E3-1245v5 CPU, 16GB of memory, and a 1Gbps ethernet interface.

\begin{figure}
  \centering
  \includegraphics[width=\linewidth]{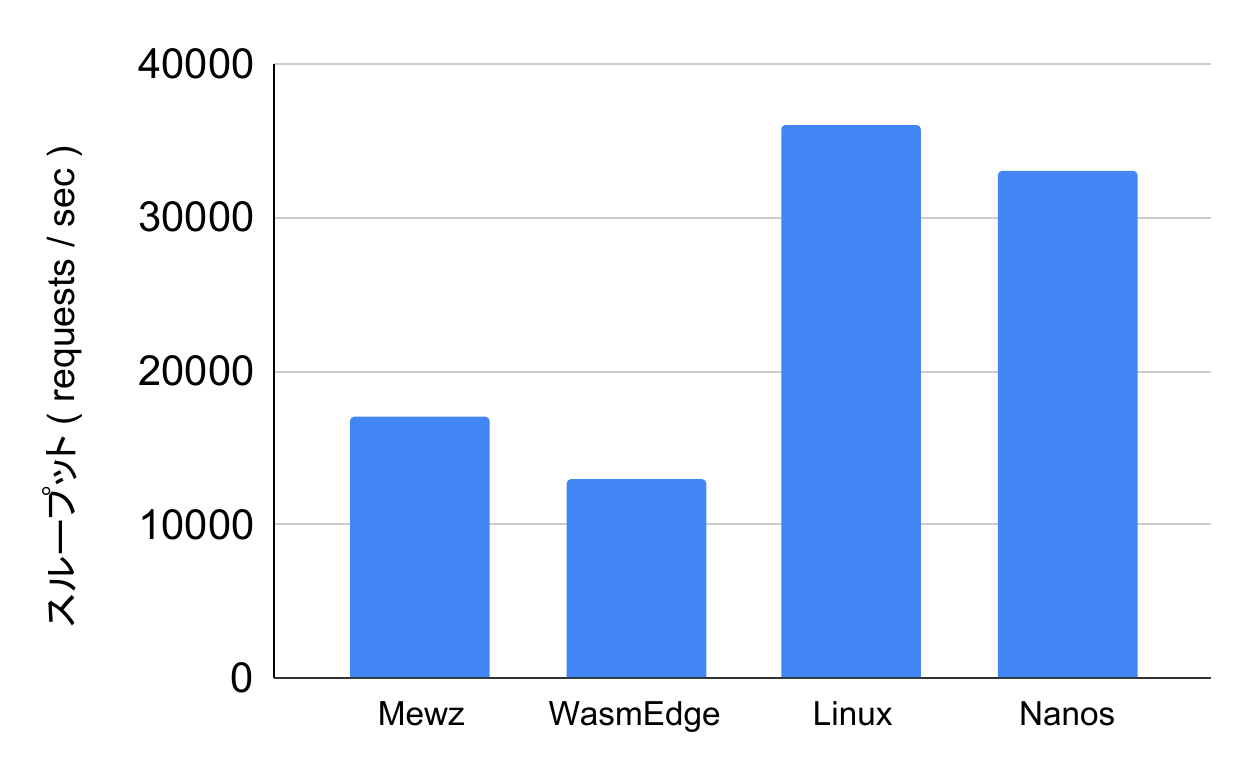}
  \caption{Result of the performance evaluation}
  \label{fig:result}
\end{figure}

The result of the evaluation is shown in the figure \ref{fig:result}.
The throughput of Mewz is 1.3 times higher than that of WasmEdge on Linu.
This result suggests that Mewz has a lower overhead than WasmEdge on Linux.
Mewz allows Wasm to call kernel functions directly, which reduces the overhead of system calls.
Additionally, Mewz entails no overhead of a Wasm runtime or other functionality that is not necessary for Wasm execution.

However, Linux and Nanos achieved 2.5 times and 2.2 times higher throughput than Mewz, respectively.
This is thought to be because the optimization of the Wasm binary generated by the Rust compiler is insufficient\cite{wasm-opt}.
In addition, Linux and Nanos have a high-performance I/O multiplexing mechanism called epoll, while WASI provides only a function called poll\_oneoff.
poll\_oneoff is equivalent to the POSIX poll function, so this has less performance than epoll.

We verified the cause of the performance difference by profiling Mewz.
We classified the source codes of Mewz into six categories and measured the CPU clock cycles spent on each processing during the benchmark.
The measurement results are shown in the table \ref{tb:profile}.
As a result, executing Wasm code and executing WASI functions occupy a large proportion.

\begin{table}[h]
  \caption{Profile of Mewz}
  \label{tb:profile}
  \centering
  \begin{tabular}{cr}
      \hline
      Classification of source codes                 & Elapsed clock cycles        \\
                            &                \\
      \hline \hline
      Executing Wasm codes      & 6,034,237,546  \\
      Executing WASI functions       & 4,952,635,522  \\
      Memory management & 9,174,456      \\
      Timer        & 4,304,220      \\
      Packet processing                & 19,006,622,184 \\
      Network driver         & 650,395,874    \\
      \hline
  \end{tabular}
\end{table}

\section{Related Work}
\label{sec:related}

Firecracker is a virtual machine monitor based on KVM\cite{firecracker}.
Firecracker was developed with the goals of fast boot times, small memory footprint, and near-physical machine performance, while ensuring a high degree of isolation.
These goals are intended for serverless computing in the public cloud.
To achieve these goals, Firecracker has a minimal set of hardware features to emulate.
This feature allows Firecracker to reduce the hypervisor device emulation overhead of virtual machines.
In fact, Firecracker achieves faster boot times and a smaller memory footprint than conventional virtual machines.
Thus, Firecracker takes the approach of reducing the overhead of the hypervisor in order to achieve lightweight virtual machines.
On the other hand, this study aims to reduce the overhead of the guest OS by using unikernel.
These approaches are not in competition with each other and can be realized independently.

gVisor is a container runtime that runs containers strongly isolated from the host system\cite{gvisor,gvisor-performance}.
gVisor was also developed to reduce the overhead of virtualization in serverless computing.
However, unlike Firecracker, gVisor does not use virtual machines for isolation, but rather sandboxes containers to ensure isolation.
gVisor implements a kernel running in user space.
This kernel intercepts system calls from the application and executes them in the user space kernel.
Since gVisor does not use a virtual machine, its startup time and memory footprint are smaller than those of a virtual machine\cite{gvisor-performance}.
However, there is a performance overhead due to hooking system calls in the application\cite{gvisor}.
Thus, gVisor endeavors to reduce the weight of the virtualization layer by increasing the isolation of containers and eliminating the need for virtual machines.
On the other hand, this study does not adopt the approach of increasing the isolation of container runtime as gVisor does.
This is because this paper aims to solve the OS and CPU architecture dependence of container images in addition to the lightweighting of the virtualization layer.

\section{Conclusion}
\label{sec:conclusion}

This paper proposed a new system that combines Wasm and unikernels.
It solves the problems of portability and overheads when using containers and VMs in cloud computing.
By adopting Wasm, it enables applications to be run on any host operating system and CPU architecture, unlike container images.
Running Wasm as a unikernel reduces the overhead of guest OS accompanying virtual machine isolation.
To implement this architecture, we developed a unikernel with WASI API and an AoT compiler that converts Wasm to native code.
We evaluated the performance of the system by running a simple HTTP server compiled into Wasm.
The result showed that it ran Wasm applications with lower overhead than an existing Wasm runtime.

\bibliographystyle{plain}
\bibliography{main}

\end{document}